\documentclass[traditabstract]{aa} 
\usepackage{graphicx}
\usepackage{txfonts}
\usepackage{natbib}
\usepackage{subfigure}
\newcommand{\kms}{\mbox{${\rm km\,s}^{-1}$}}
\newcommand{\kmsa}{\mbox{${\rm km\,s}^{-1}\,a^{1/3}$}}
\newcommand{\Msolar}{\mbox{${M}_{\sun}$}}
\newcommand{\Rsolar}{\mbox{${R}_{\sun}$}}

\newcommand{\Mjup}{\mbox{${M}_{J}$}}
\newcommand{\Mearth}{\mbox{${M}_{\oplus}$}}
\newcommand{\rhosun}{\mbox{$\rho_{\sun}$}}
\newcommand{\Rjup}{\mbox{${R}_{J}$}}
\newcommand{\rhojup}{\mbox{$\rho_{J}$}}

\newcommand{\teff}{\mbox{$T_{\rm eff}$}}
\newcommand{\logg}{\mbox{$\log g$}}
\newcommand{\vsini}{\mbox{$v \sin i_{\ast}$}}
\newcommand{\mictrb}{\mbox{$\xi_{\rm t}$}}
\newcommand{\mactrb}{\mbox{$v_{\rm mac}$}}

\newcommand{\halpha}{\mbox{$H_\alpha$}}

\newcommand\T{\rule{0pt}{2.2ex}}

\begin{document}
   \title{WASP-42~b and WASP-49~b: two new transiting sub-Jupiters. \thanks{Based on photometric observations made with WASP-South, EulerCam on Euler-Swiss telescope, 
the Belgian TRAPPIST telescope, the Faulkes South Telescope and spectroscopic observations obtained with CORALIE on the Euler-Swiss telescope
and HARPS on the ESO 3.6~m telescope (Prog. ID: 087.C-0649)}$^{,}$\thanks{The photometric time series and radial velocity data in this work are only available in electronic form
at the CDS via anonymous ftp to cdsarc.u-strasbg.fr (130.79.128.5) or via http://cdsweb.u-strasbg.fr/cgi-bin/qcat?J/A+A/
}}

   \subtitle{}

   \author{M.~Lendl
          \inst{1}
          \and
          D.~R.~Anderson
          \inst{2}
          \and 
          A.~Collier-Cameron
          \inst{3}
          \and        
          A.~P.~Doyle
          \inst{2}
          \and
          M.~Gillon
          \inst{4}
          \and
          C.~Hellier
          \inst{2}
          \and
          E.~Jehin
          \inst{4}
          \and
          T.~A.~Lister
          \inst{5}
          \and
          P.~F.~L.~Maxted
          \inst{2}
          \and
          F.~Pepe
          \inst{1}
          \and
          D.~Pollacco
          \inst{6}
          \and
          D.~Queloz
          \inst{1}
          \and          
          B.~Smalley
          \inst{2}
          \and
          D.~S\'egransan
          \inst{1}
          \and
          A.~M.~S.~Smith
          \inst{2}
          \and
          A.~H.~M.~J.~Triaud
          \inst{1}
          \and
          S.~Udry
          \inst{1}
          \and
          R.~G.~West
          \inst{7}
          \and
          P.~J.~Wheatley
          \inst{8}
          }

   \institute{Observatoire de Gen\`eve, Universit\'e de Gen\`eve, Chemin des maillettes 51, 1290 Sauverny, Switzerland
              \email{monika.lendl@unige.ch}
         \and 
             Astrophysics Group, Keele University, Staffordshire, ST5 5BG, United Kingdom
         \and
             School of Physics and Astronomy, University of St. Andrews, North Haugh, Fife, KY16 9SS, United Kingdom
         \and
             Universit\'e de Li\`ege, All\'ee du 6 ao\^ut 17, Sart Tilman, Li\`ege 1, Belgium
         \and
             Las Cumbres Observatory, 6740 Cortona Drive Suite 102, Goleta, CA 93117, USA
         \and             
             Astrophysics Research Centre, School of Mathematics \& Physics, Queen's University, University Road, Belfast BT7 1NN, United Kingdom
         \and
             Department of Physics and Astronomy, University of Leicester, Leicester, LE1 7RH, United Kingdom            
         \and 
             Department of Physics, University of Warwick, Coventry CV4 7AL, United Kingdom
             }

   \date{}
 
  \abstract{
We report the discovery of two new transiting planets from the WASP survey. WASP-42~b is a $0.500 \pm 0.035$ {\Mjup} planet orbiting
a K1 star at a separation of $0.0548 \pm 0.0017$~AU with a period of $4.9816872\pm7.3\times10^{-6}$ days. The radius of WASP-42~b is 
$1.080 \pm 0.057$~{\Rjup} while its equilibrium temperature is $T_{eq}=995\pm 34$~K. We detect some evidence for a small but non-zero eccentricity
of $e=0.060\pm0.013$.
WASP-49~b is a $0.378\pm0.027$~{\Mjup} planet around an old G6 star. It has a period of $2.7817387\pm5.6\times10^{-6}$ days and a 
separation of $0.0379\pm0.0011$ AU. This planet is slightly bloated, having a radius of $1.115\pm0.047$~{\Rjup} and an 
equilibrium temperature of $T_{eq}=1369\pm39$~K.
Both planets have been followed up photometrically, and in total we have obtained 5 full and one partial transit light curves of
WASP-42 and 4 full and one partial light curves of WASP-49 using the Euler-Swiss, TRAPPIST and Faulkes South telescopes.  
}


   \maketitle
%

\section{Introduction}

Since the discovery of the first extrasolar planet around a Solar-type star by \citet{Mayor95} over 700 exoplanets have been found.
In recent years an increasingly large number of planets have been discovered
by transit surveys, i.e. surveys that search for planets that pass in front of their host stars producing characteristic photometric signals.
At the time of writing, the two most prominent ground-based transit surveys are HATnet \citep{Bakos04} and WASP \citep{Pollacco06}.

From ground-based transit surveys, a population of hot close-in giant planets has been explored. These Hot Jupiters are located at small separations,
having periods of typically 2 to 5 days. While the formation of 
gas giants is accepted to take place in cooler regions of the protoplanetary disc at several AU from the star, the process
by which the planets have migrated to the proximity of the star has been a matter of debate. 
Migration can occur from interactions with the disc itself \citep{Goldreich80,Lin96}, or due to dynamical interactions between massive bodies in the system, namely 
scattering between planets \citep{Rasio96} and Kozai migration \citep{Kozai62,Eggleton01,Wu03}, which put the planet on a highly eccentric orbit that then becomes
tidally circularized. Recent measurements of the sky-projected spin-orbit angles have revealed several misaligned systems \citep{Hebrard08,Triaud10,Winn10}, a fact 
which could be caused by planets undergoing Kozai migration \citep{Fabrycky07,Wu07,Triaud10}.

Many of these planets have been shown to posses radii which are larger than expected from models of irradiated planets \citep[e.g.][]{Fortney07}. Low densities appear
to be common for Saturn-mass planets (e.g. WASP-39~b, \citealt{Faedi11}) and up to planets of the mass of Jupiter (e.g. \mbox{HAT-P-32 b}, \citealt{Hartman11}). The most striking
examples are WASP-17~b having a density of $\rho=0.06~{\rhojup}$ \citep{Anderson10a,Anderson11a} and \mbox{Kepler-12 b} with a density of $\rho=0.09~{\rhojup}$ \citep{Fortney11}.
It appears that there is a mechanism depositing energy into the planet and thus inflating it or slowing the contraction of the planet since its formation.
A variety of mechanisms have been proposed to account for this effect: the deposition of kinetic energy stemming from strong winds driven by large day/night
temperature contrasts \citep{Showman02}; enhanced opacities due to higher planetary metallicity causing a slow-down in contraction 
\citep{Burrows07}; heating by tidal forces from the circularization, synchronization and re-alignment of the planetary orbit \citep{Bodenheimer01}; reduced heat transport 
efficiency by layered convection inside the planet \citep{Chabrier07}, and Ohmic heating from currents induced through winds in the planetary atmosphere 
\citep{Batygin10}. It has been found that the degree to which the planet is bloated is correlated with the incident stellar flux \citep{Demory11,Enoch11,Enoch12,Laughlin11}, 
favouring models which incorporate the absorption of stellar radiation as a cause of the planetary bloating.

In this paper, we announce the discovery of two additional transiting planets from the WASP-South survey. WASP-42~b is a 
~0.5 {\Mjup} planet in a ~5 day orbit around a K1 star, and WASP-49~b, a bloated ~0.4 {\Mjup} planet, is orbiting a metal poor G6 star every 2.8 days.
In Section \ref{sec:obs} we present the discovery and follow-up observations of WASP-42 and WASP-49, leading to their identification
as harbouring transiting extrasolar planets. In Section \ref{sec:detsys} we describe the analysis of our data before putting the two
planets into context in Section \ref{sec:disc}.

\section{Observations}
\label{sec:obs}

\begin{figure}[t]
 \centering
 \includegraphics[width=\linewidth]{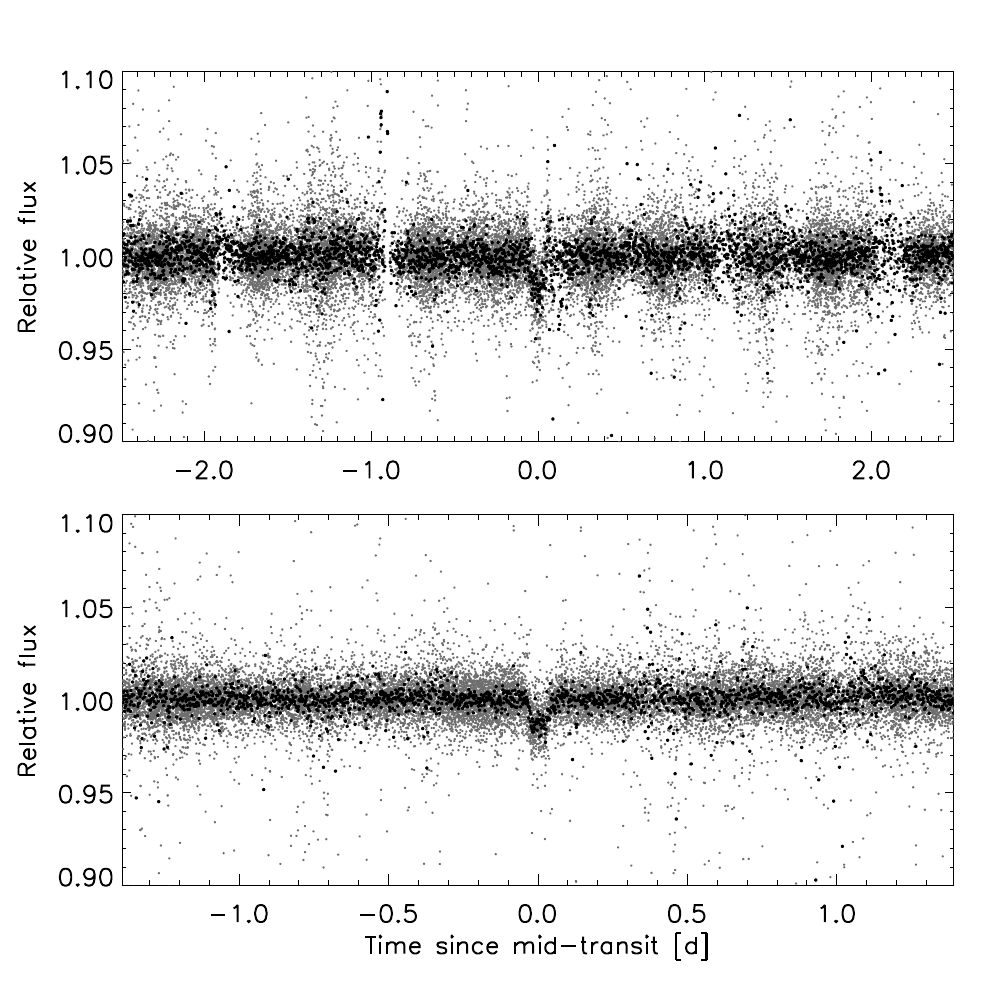}
    \caption{Phase-folded WASP photometry of WASP-42 (top) and WASP-49 (bottom). The unbinned data are depicted in gray; the black points represent the same data set binned to 2 minutes.}
    \label{fig:W4249SW}
\end{figure}

  \subsection{WASP-Photometry}

The objects WASP-42 (\mbox{2MASS 12515557--4204249}) and WASP-49 (\mbox{2MASS 06042146--1657550}) were observed using the WASP survey telescopes. 
The WASP survey is operated from two sites, one in each hemisphere: the Observatorio del Roque de los Muchachos in the Canary Islands
in the North, and the Sutherland Station of the South African Astronomical Observatory (SAAO) in the South. Each site is equipped with eight
commercial 11~cm, f=200~mm Canon lenses on a single mount. Details of the WASP survey,
its photometric reduction and candidate selection process can be found in \citet{Pollacco06} and \citet{Cameron07}. Both targets presented
in this work were observed exclusively from the southern WASP site.
In total 25880 data points were obtained for WASP-42~between May 2006 and April 2008, while 18461 data points were obtained for  WASP-49~between October 2006 and March 2010. The data obtained for both targets are shown in Figure \ref{fig:W4249SW}. In both cases  
a periodically occurring transit-like signal was detected in the data using the transit-search 
algorithms of \citet{Cameron06}, triggering a closer inspection of the targets and their selection for further study.

\subsection{Spectroscopic Observations}

\begin{figure}[t!]
 \centering
 \includegraphics[width=0.9\linewidth]{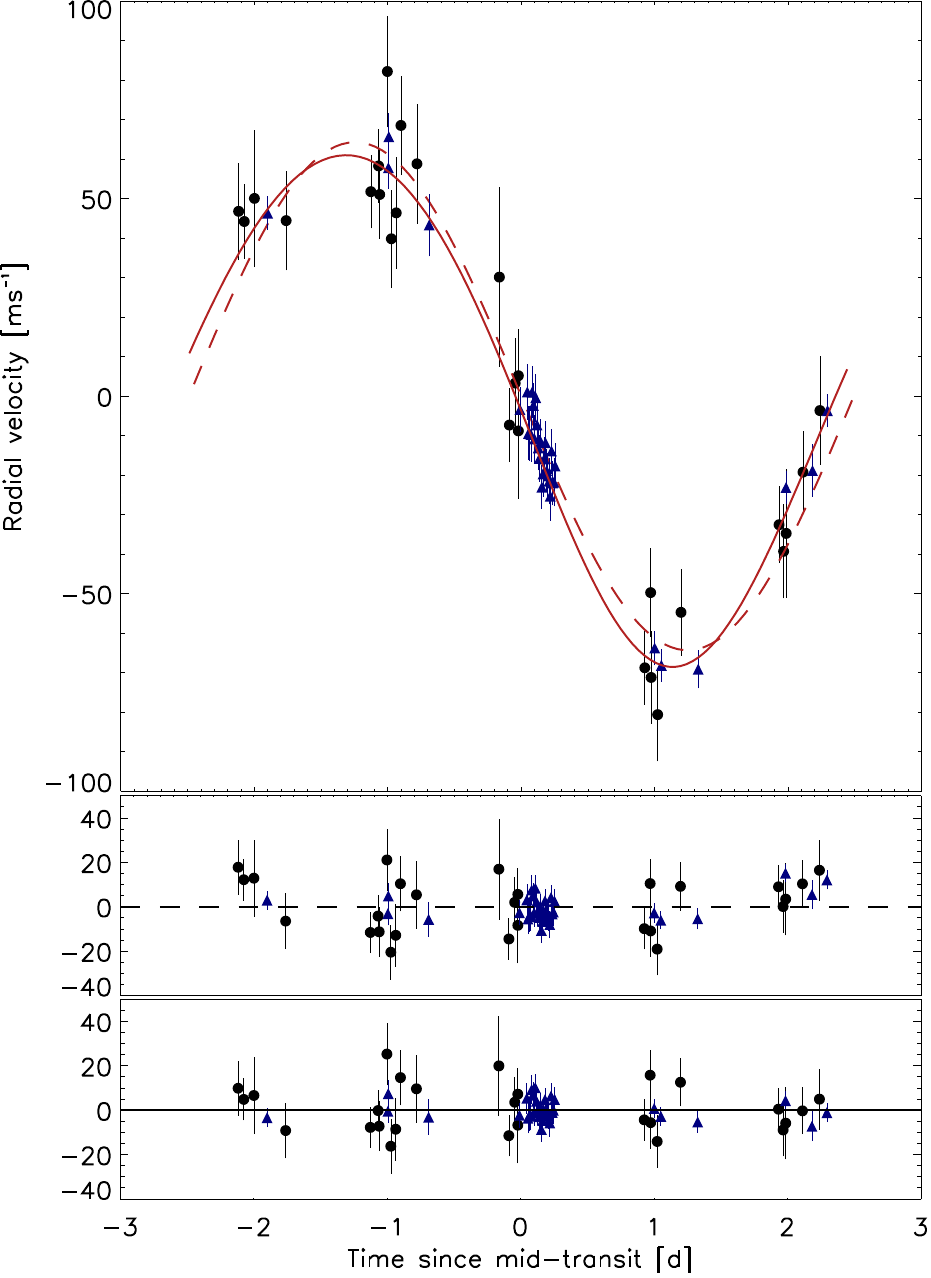}
    \caption{Upper panel: Radial Velocity observations of WASP-42 together with the circular (dashed line) and eccentric (solid line) models. Black dots denote CORALIE data,
while blue triangles denote HARPS data. Middle panel: residuals of the circular model. Lower panel: residuals of the eccentric model.}
    \label{fig:W42rv}
\end{figure}

\begin{figure}[h!]
 \centering
 \includegraphics[width=0.9\linewidth]{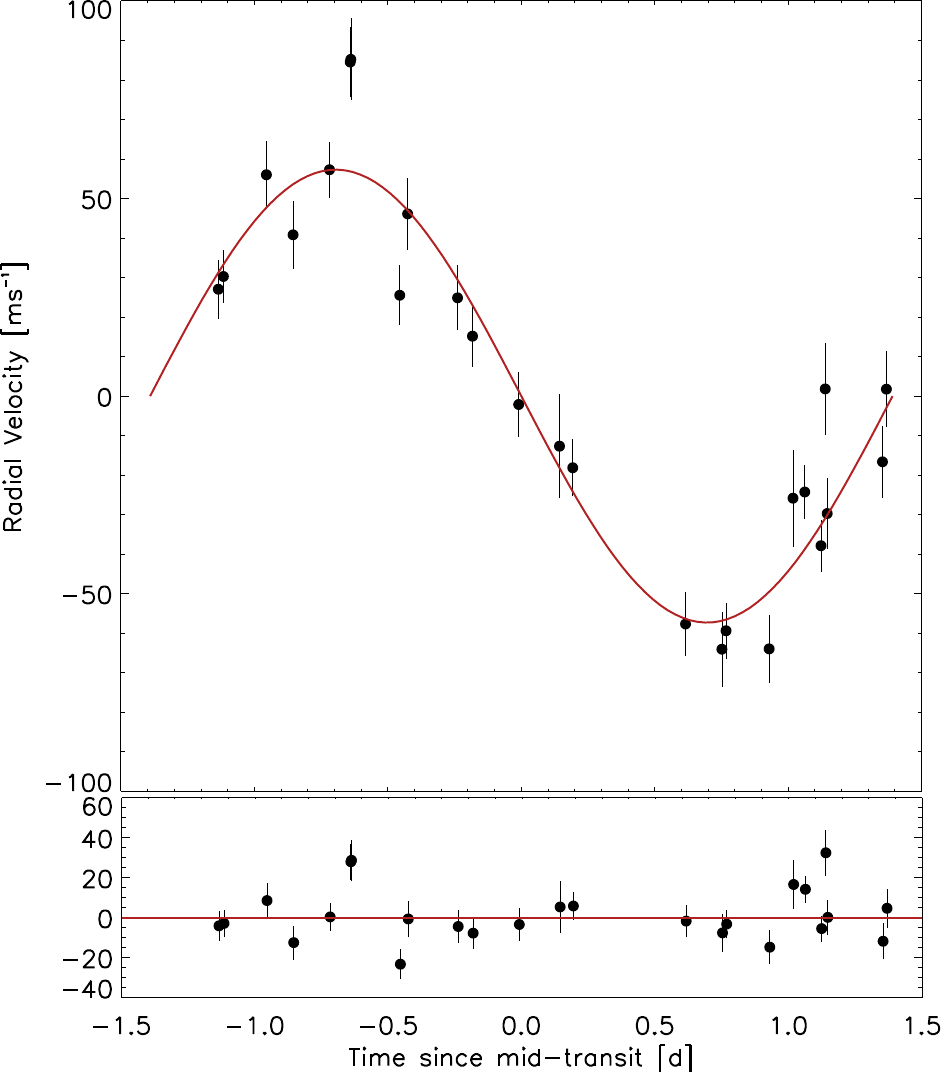}
    \caption{Radial Velocity observations of WASP-49 together with the respective model and residuals. All data have been obtained with CORALIE.}
    \label{fig:W49rv}
\end{figure}
 
\begin{figure*}
 \subfigure[\label{fig:W42bis}Bisector measurements of WASP-42 against the measured radial velocity. Black dots denote 
data points from CORALIE while blue triangles denote points obtained with HARPS.]{
 \includegraphics[width=0.45\linewidth]{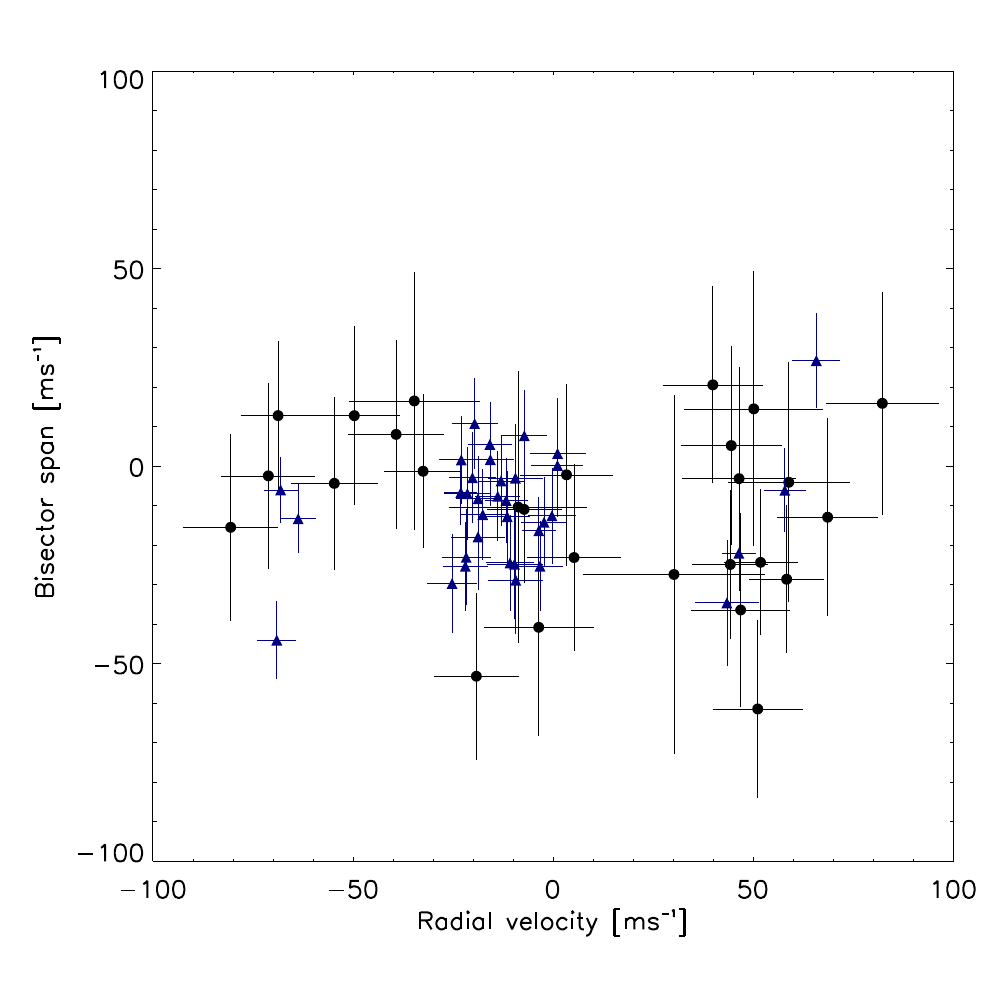}
}
\hspace{30pt}
 \subfigure[\label{fig:W49bis}Bisector measurements of WASP-49 against the measured radial velocity. All data have been obtained with CORALIE.]{
 \includegraphics[width=0.45\linewidth]{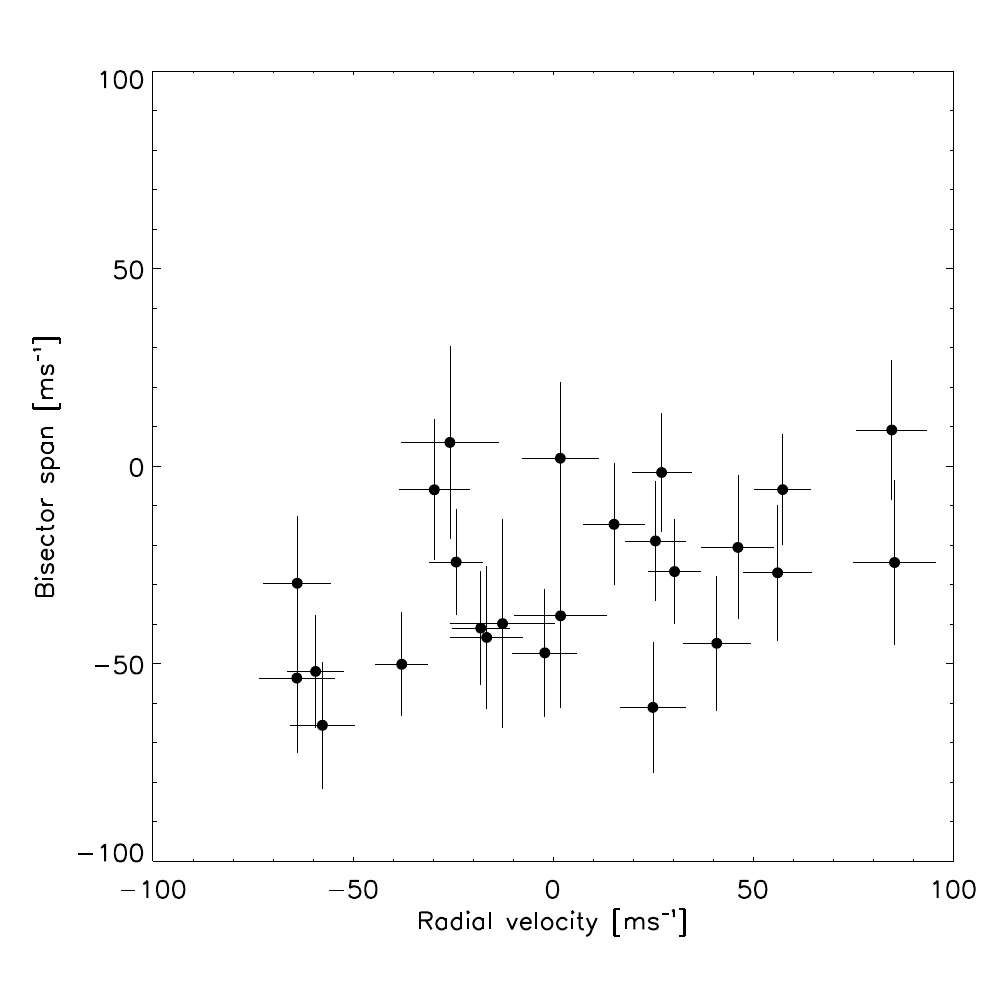}
}
\caption{\label{fig:W4249bis}Bisector spans measured for WASP-42 \textit{(left)} and WASP-49 \textit{(right)}.}
\end{figure*}

WASP-42 and WASP-49 were observed with the CORALIE spectrograph mounted on the 1.2~m Euler-Swiss telescope at the La Silla site (Chile). For WASP-42 and WASP-49, 27 and 25 data points were
obtained from April 2010 to March 2011, and from August 2009 to October 2011, respectively. For all spectroscopic observations we obtained radial velocities using the 
weighted Cross-Correlation technique \citep{Baranne96} as implemented in the CORALIE and HARPS reduction pipelines. The radial velocities folded on the transit ephemeris 
are shown in Figures \ref{fig:W42rv} and \ref{fig:W49rv}. In order to ensure that the radial-velocity variations are not 
caused by star spots we checked the CCF bisector spans (shown in Figure \ref{fig:W4249bis}) according to the method described by \citet{Queloz01}. After the planetary nature of the signal had been confirmed, 
WASP-42 was also observed with the HARPS spectrograph \citep{Mayor03} which is located at the ESO 3.6~m telescope at La Silla observatory. 
35 data points were obtained with HARPS, 25 of which were observed nearly consecutively on 4 April 2011 (UT) in order to measure the Rossiter--McLaughlin effect 
\citep{Rossiter24,McLaughlin24} and thus determine the projected spin--orbit angle of the WASP-42 system. However, due to an inaccurate ephemeris at the time of planning of these 
observations, the continuous observations began only during the egress of the planet (see Figure \ref{fig:W42ros}) and thus the Rossiter--McLaughlin effect was not measured.

  \subsection{Follow-up Photometry}

\begin{figure*}
 \centering
 \includegraphics[width=\linewidth]{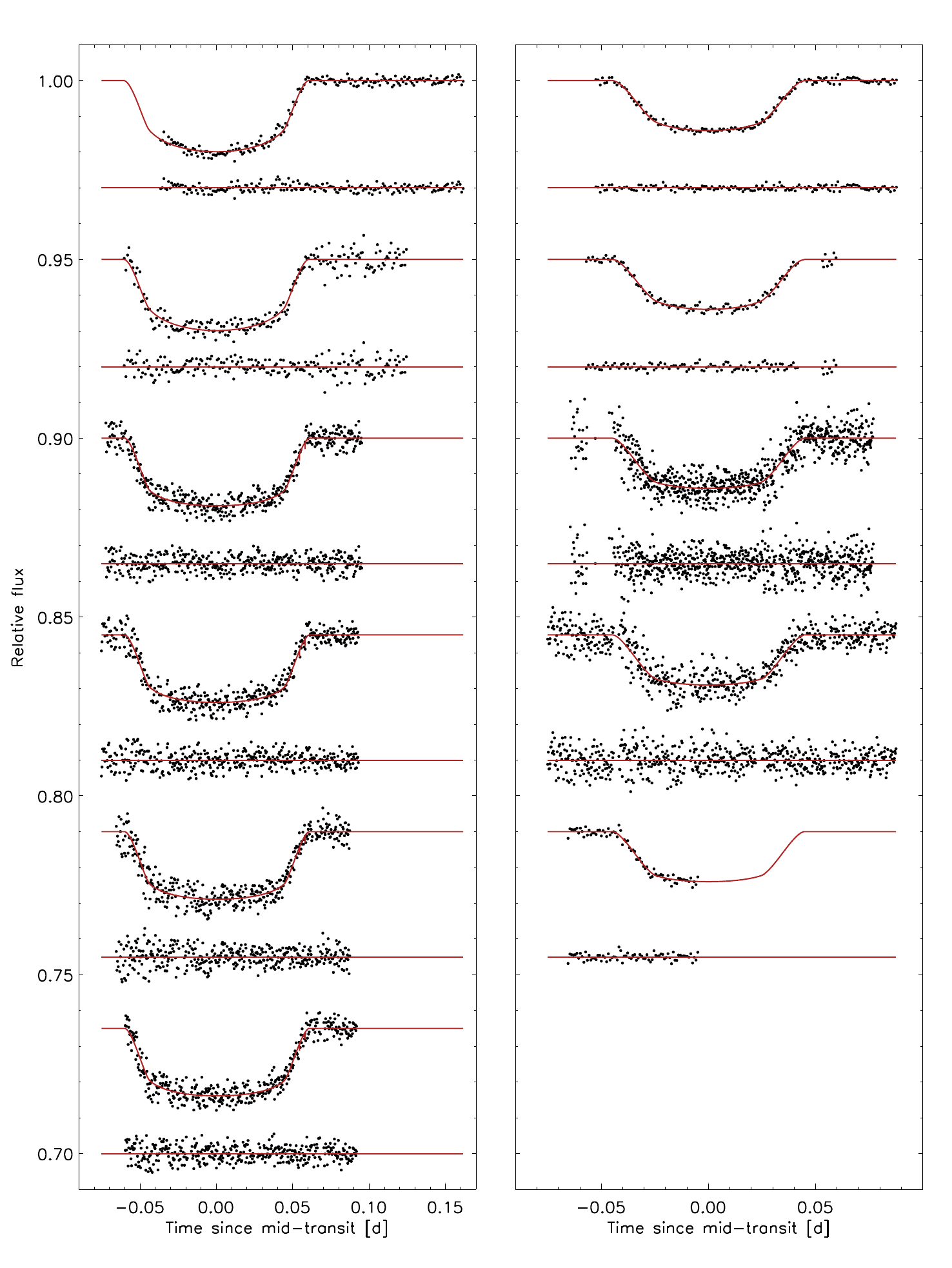}
    \caption{Follow-up photometry of WASP-42 \textit{(left)} and WASP-49 \textit{(right)} together with the respective model and residuals. 
Each light curve has been divided by the respective photometric baseline model as described in Section \ref{sec:comban}. The light curves of WASP-42 are (from top to bottom) 
EulerCam on March 20 and 25 2011, and TRAPPIST on March 5, 10, 30 and April 4 2011 (UT). The light curves of WASP-49 are (from top to bottom) 
EulerCam on January 19 and March 24 2011, TRAPPIST on January 19 and 24 October 2011, and FTS on 21 March 2011 (UT).}
    \label{fig:W4249phot}
\end{figure*}

Two transits each of both WASP-42~b and WASP-49~b were observed using EulerCam between January and March 2011. EulerCam is an \emph{e2v} 4k x 4k back-illuminated
deep-depletion silicon CCD detector which was installed at the Cassegrain focus of the 1.2~m Euler-Swiss telescope in September 2010. The field of view of EulerCam is 
15.68 x 15.73 arcmin, producing a resolution of 0.23 arcsec per pixel. The instrument is operated at a temperature of --115$\,^{\circ}\mathrm{C}$, 
cool enough to make dark current negligible (measured at less than $2.6~e^{-}$ in 30 minutes). EulerCam can be read out using either one or four ports, 
giving readout times of 6.5~s and 25~s. For single-port readout the readout noise is $\sim5~e^{-}$/pixel. For four-port readout the readout noise is the same except that the lower-left port shows a slightly elevated noise of $\sim 8~e^{-}$/pixel. As flat-field uncertainties 
are often a limiting factor in high-precision photometry, EulerCam uses a feedback scheme for the telescope guiding
to keep the stars on the same locations on the detector during the observations. This system, dubbed ``Absolute Tracking'' is based on a combination of the 
\emph{SCAMP} \citep{Bertin06} and \emph{Sextractor} \citep{Bertin96} software packages. After recording an image, the positions of the stars on the image are extracted and then 
matched with a catalogue. From this match, the offset of the telescope from the nominal position is calculated using a PID algorithm and the telescope pointing is adjusted between exposures
in order to compensate for drifts. Due to technical problems the 4-port readout mode was not available during the first transit of WASP-49, leading to a slightly degraded 
time sampling. All EulerCam observations were done with a defocus of $0.1$~mm in order to improve the duty cycle and spread the light over more pixels and thereby
improve the sampling of the PSF.

Four transits of WASP-42~b and two transits of WASP-49~b were observed with the automated Belgian TRAPPIST telescope, also located at La Silla. For details
of TRAPPIST see \citet{Gillon11a} and \citet{Jehin11}. Again, the telescope was defocused giving a FWHM of 3.2 arcsec on the images. 

Finally, one partial transit of WASP-49~b was observed with the Faulkes Telescope South (FTS) which is based at Siding Spring, Australia.
Details of all observations are summarized in Table \ref{tab:phot} and the light curves are depicted in Figure \ref{fig:W4249phot}. 

All follow-up light curves were obtained from bias- and flat-field-corrected images using relative Aperture Photometry where several apertures were tested and reference stars
were chosen with great care. In the case of TRAPPIST, IRAF \footnote[1]{IRAF is distributed by the National Optical Astronomy Observatories, which are operated by the 
Association of Universities for Research in Astronomy, Inc., under cooperative agreement with the National Science Foundation.} was used in the reduction process. 
For EulerCam, the standard reduction procedure is as follows. After overscan, bias and flat-field correction, the photometry is extracted from the images for 
several circular apertures with radii ranging from 12 to 50 pixels placed on the target and all other bright stars in the field. 
The optimal combination of reference stars is found by using the target itself to measure the quality of the references. We start the process with the reference 
star with which we obtain the lowest RMS transit light curve and iteratively add those other references which yield the best improvement at any given step. 
Typically, the final reference sources is made of between three and ten stable stars of similar brightness and color to the target and which are not affected by any 
short-term variations (caused, e.g., by proximity to bad pixels).

During the observations of WASP-49 with EulerCam we noticed a faint source, 9.2 arcsec North of the target. It is outside the smaller apertures for which we extracted
photometry in the analysis of the EulerCam data. As the transit is present in light curves created using these apertures (while the data are noisier than for
larger apertures), we are certain the transit is on the main target. However, in order to evaluate the contamination for larger apertures,
we re-observed WASP-49 on 15 January 2012 with EulerCam without defocusing the telescope. The observations were performed outside of transit, a total of 8 images were obtained 
over 10 minutes using an r' Gunn filter. We find the brightness ratio between the target and the contaminant to be $1:0.00343\pm (5\times10^{-5})$, 
and, as its impact on the observed transit depth is much smaller ($4\times10^{-5}$) than the 1-$\sigma$-error quoted in Table \ref{tab:par}, we neglect it in the analysis presented below.

\begin{table*}
\centering                        
\begin{tabular}{c c c c c c c}       
\hline\hline                
  target & date (UT) & telescope/instrument & Filter & $T_{c}$ [$HJD_{TBD} - 2450000$] & $\beta_{red}$\tablefootmark{1}& RMS [relative flux, per 2 min] \T \\   
\hline                   
   WASP-42 & 05-03-2011 & TRAPPIST & I+z' & $5625.65821\pm0.00044$ & 1.13 & $1.2\times10^{-3}$ \T \\       
   WASP-42 & 10-03-2011 & TRAPPIST & I+z' & $5630.64098\pm0.00055$ & 1.45 & $1.4\times10^{-3}$ \\   
   WASP-42 & 20-03-2011 & EulerCam & r' Gunn & $5640.60221\pm0.0040$ & 1.43 & $0.9\times10^{-3}$ \\
   WASP-42 & 25-03-2011 & EulerCam & r' Gunn & $5645.58600\pm0.00057$ & 1.06 & $1.9\times10^{-3}$ \\
   WASP-42 & 30-03-2011 & TRAPPIST & I+z' & $5650.56771\pm0.00066$ & 1.18 & $1.4\times10^{-3}$ \\   
   WASP-42 & 04-04-2011 & TRAPPIST & I+z' & $5655.54921\pm0.00056$ & 1.15 & $1.2\times10^{-3}$ \\   
   WASP-49 & 19-01-2011 & EulerCam & r' Gunn & $5580.59412\pm0.00041$ & 1.26 & $0.5\times10^{-3}$ \\  
   WASP-49 & 19-01-2011 & TRAPPIST & I+z' & $5580.59523\pm0.00074$ & 1.00 & $1.5\times10^{-3}$ \\   
   WASP-49 & 21-02-2011 & FTS & z' Gunn & $5613.976\pm0.011$ & 1.00 & $0.7\times10^{-3}$ \\   
   WASP-49 & 24-03-2011 & EulerCam & r' Gunn & $5644.57521\pm0.00035$ & 1.00 & $0.7\times10^{-3}$ \\
   WASP-49 & 24-10-2011 & TRAPPIST & I+z' & $5858.76832\pm0.00048$ & 1.08 & $1.4\times10^{-3}$ \\   
\hline                                  
\end{tabular}
\caption{\label{tab:phot}Summary of follow-up photometry. Target, date, telescope and filter are given for each observation together with 
the mid-transit time, red noise amplitude and the RMS of the binned (2 minutes) residuals. \newline \tablefoottext{1}{as defined in \citet{Winn10}}}
\end{table*}

\begin{figure}
 \centering
 \includegraphics[width=\linewidth]{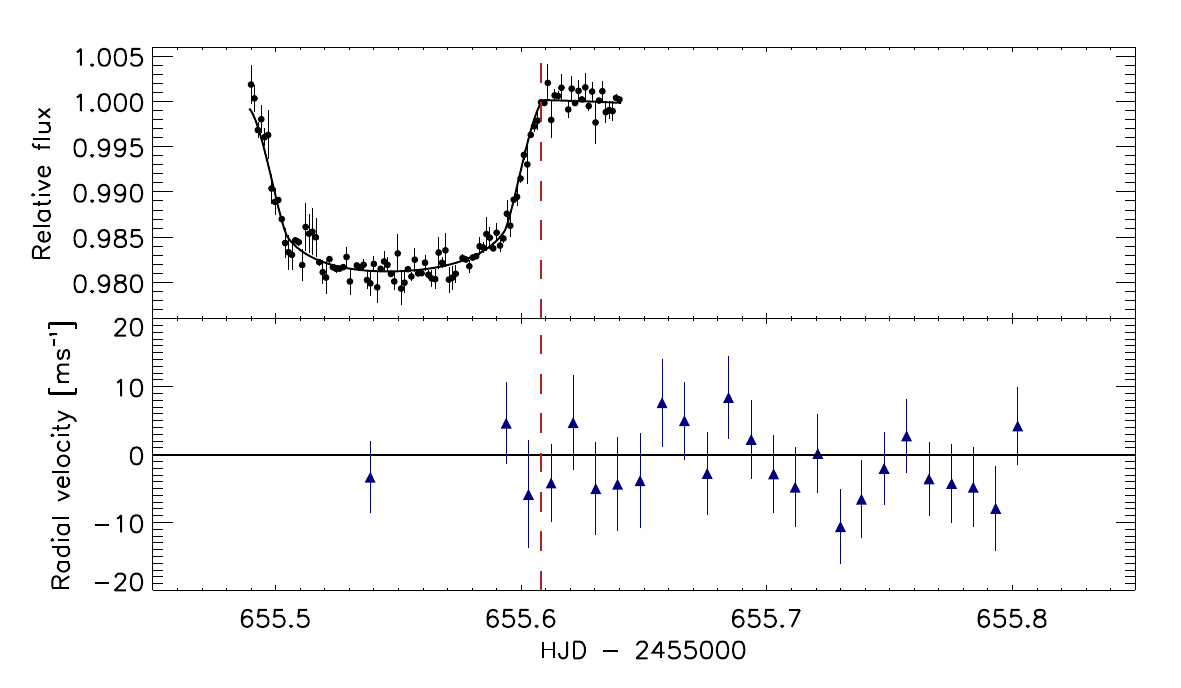}
    \caption{Photometric and radial velocity observations obtained on April 4 2011 (UT). Upper panel: TRAPPIST transit light curve binned in 2 minute intervals.
Lower panel: the the HARPS time series observation aiming at measuring the Rossiter--McLaughlin effect. Depicted
are the residual values from a purely Keplerian model. The dashed red line indicates the time of fourth contact.}
    \label{fig:W42ros}
\end{figure}

\section{Determination of System Parameters}
\label{sec:detsys}

\subsection{Stellar Parameters}
\label{sec:spar}

\subsubsection{Rotation period}

 We have analyzed the WASP lightcurves of WASP-42 and WASP-49 to determine
whether they show periodic modulation due to the combination of magnetic
activity and the rotation of the star. 
We used the sine-wave fitting method described in \citet{Maxted11}
to calculate periodograms for both stars over 4096 uniformly spaced
frequencies from 0 to 2.5 cycles/day. The false alarm probability (FAP) for
the strongest peak in these periodograms was calculated using the boot-strap
Monte Carlo method also described in \citet{Maxted11}. Variability due to
star spots is not expected to be coherent on long timescales as a consequence
of the finite lifetime of star-spots and differential rotation in the
photosphere, and so for both stars we analyzed the data from each observing
season independently. 
 
 We did not find any significant periodic signals (FAP$< 0.05$) in our data
apart from frequencies near 1\,cycle/day and its harmonics. 
We examined the distribution of amplitudes
for the most significant frequency in each Monte Carlo trial and used  these
results to estimate a 95\,\% upper confidence limit of 1\,milli-magnitude for
the amplitude of any periodic signal in the lightcurves for both WASP-42 and WASP-49. 

\subsubsection{Spectroscopic Analysis}

For each star a total of 11 (WASP-42) and 23 (WASP-49) individual CORALIE spectra were co-added to produce
single spectra with typical S/N values of 60:1 (WASP-42) and 100:1 (WASP-49). The standard pipeline
reduction products were used in the analysis.

The analyses were performed using the methods given in
\citet{Gillon2009a}. The \halpha\ line was used to determine the
effective temperature (\teff), while the Na {\sc i} D and Mg {\sc i} b lines
were used as surface gravity (\logg) diagnostics. The parameters obtained from
the analysis are listed in Table \ref{tab:spar}. The elemental abundances
were determined from equivalent width measurements of several clean and
unblended lines. Values for microturbulence (\mictrb) were determined from
Fe~{\sc i} using the method of \citet{Magain84}. The quoted error
estimates include those given by the uncertainties in \teff, \logg\ and \mictrb,
as well as the scatter due to measurement and atomic data uncertainties.

The projected stellar rotation velocities (\vsini) were determined by fitting the
profiles of several unblended Fe~{\sc i} lines. Values for macroturbulence
(\mactrb) of 1.4 $\pm$ 0.3 {\kms} (WASP-42) and 2.9 $\pm$ 0.3 {\kms} (WASP-49) were 
assumed, based on the tabulation by \citet{Gray08}. 
For both cases, an instrumental FWHM of 0.11 $\pm$ 0.01~{\AA} was determined from the telluric 
lines around 6300\AA. Best fitting values of
\vsini\ = 2.7 $\pm$ 0.4 ~\kms\ (WASP-42) and \vsini\ = 0.9 $\pm$ 0.3 ~\kms\ (WASP-49) were obtained.

There is no significant detection of lithium in the spectra of either star, with equivalent
width upper limits of 4~{m\AA}, corresponding to an abundance upper limit of log
A(Li) $<$ 0.5 $\pm$ 0.2 (WASP-42) and 8m\AA, corresponding to an abundance upper limit of log
A(Li) $<$ 0.7 $\pm$ 0.1 (WASP-49).

\begin{table}[h]
\centering                        
\begin{tabular}{ccc} \hline \hline
Parameter  & WASP-42 & WASP-49 \T \\ \hline
RA         & $12^{\mathrm{h}}51^{\mathrm{m}}55.62^{\mathrm{s}}$& $06^{\mathrm{h}}04^{\mathrm{m}}21.47^{\mathrm{s}}$ \T \\
DEC        & $-42^{\circ}04'25.2''$ &  $-16^{\circ}57'55.1''$ \\
V mag      & 12.57 & 11.36  \\ \hline
\teff      & 5200 $\pm$ 150 K  &   5600 $\pm$ 150 K \T\\
\logg      & 4.5 $\pm$ 0.1 &   4.5 $\pm$ 0.1 \\
\mictrb    & 0.8 $\pm$ 0.2 \kms &   0.9 $\pm$ 0.2 \kms \\
\vsini     & 2.7 $\pm$ 0.4 \kms &   0.9 $\pm$ 0.3 \kms \\
{[Fe/H]}   &   0.05 $\pm$ 0.13  &$-$0.23 $\pm$ 0.07 \\
{[Na/H]}   &   0.23 $\pm$ 0.10  &$-$0.10 $\pm$ 0.08 \\
{[Mg/H]}   &   0.21 $\pm$ 0.10  &$-$0.09 $\pm$ 0.06 \\
{[Si/H]}   &   0.18 $\pm$ 0.06  &$-$0.04 $\pm$ 0.05 \\
{[Ca/H]}   &   0.11 $\pm$ 0.12  &$-$0.06 $\pm$ 0.11 \\
{[Sc/H]}   &   0.13 $\pm$ 0.09  &   0.06 $\pm$ 0.05 \\
{[Ti/H]}   &   0.27 $\pm$ 0.15  &   0.05 $\pm$ 0.06 \\
{[V/H]}    &   0.59 $\pm$ 0.13  &   0.02 $\pm$ 0.16 \\
{[Cr/H]}   &   0.12 $\pm$ 0.12  &$-$0.16 $\pm$ 0.03 \\
{[Co/H]}   &   0.30 $\pm$ 0.06  &$-$0.03 $\pm$ 0.09 \\
{[Ni/H]}   &   0.14 $\pm$ 0.09  &$-$0.14 $\pm$ 0.06 \\
log A(Li)  &   $<$ 0.5 $\pm$ 0.2 &   $<$ 0.7 $\pm$ 0.1 \\
Mass       &   0.89 $\pm$ 0.08 $M_{\sun}$ &   0.94 $\pm$ 0.07 $M_{\sun}$ \\
Radius     &   0.87 $\pm$ 0.11 $R_{\sun}$ &   0.90 $\pm$ 0.11 $R_{\sun}$ \\
Sp. Type   &   K1 &   G6 \\
Distance   &   160 $\pm$ 40 pc & 170 $\pm$ 20 pc \\ \hline
\end{tabular}
\caption{Stellar parameters of WASP-42 and WASP-49 from Spectroscopic Analysis.
\newline{\bf Note:} Mass and Radius estimates using the
\cite{Torres10} calibration. Spectral Type estimated from \teff\
using the table in \cite{Gray08}.
}
\label{tab:spar}
\end{table}

\begin{figure*}
 \subfigure[\label{fig:W42iso}Modified Hertzsprung--Russell diagram showing the location of WASP-42 together with isochrones for ages 
of (from bottom to top) 0.5, 1.0, 2.0, 3.0, 5.0, 10.0 and 13.0 Gyr. The red dashed lines indicate evolutionary paths 
for (from left to right) 1.0, 0.9 and 0.8 {\Msolar}. Isochrones and evolutionary paths have been interpolated from 
\citet{Marigo08} using $z=0.021$.]{
 \includegraphics[width=0.45\linewidth]{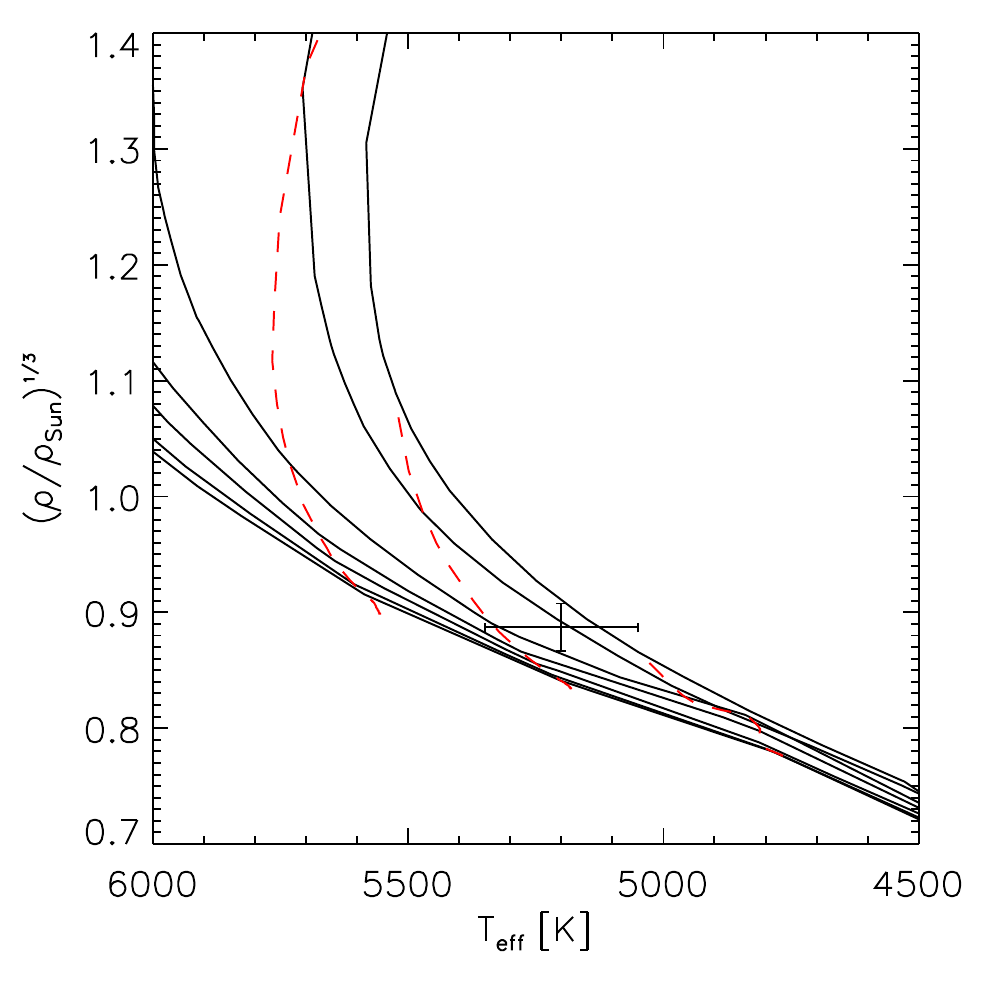}
}
\hspace{30pt}
 \subfigure[\label{fig:W49iso}Modified Hertzsprung--Russell diagram showing the location of WASP-49 together with isochrones for ages 
of (from bottom to top) 0.5, 1.0, 2.0, 3.0, 5.0, 10.0 and 13.0 Gyr. The red dashed lines indicate evolutionary paths 
for (from left to right) 1.1, 1.0, 0.9 and 0.8 {\Msolar}. Isochrones and evolutionary paths have been interpolated from 
\citet{Marigo08} using $z=0.011$.]{
 \includegraphics[width=0.45\linewidth]{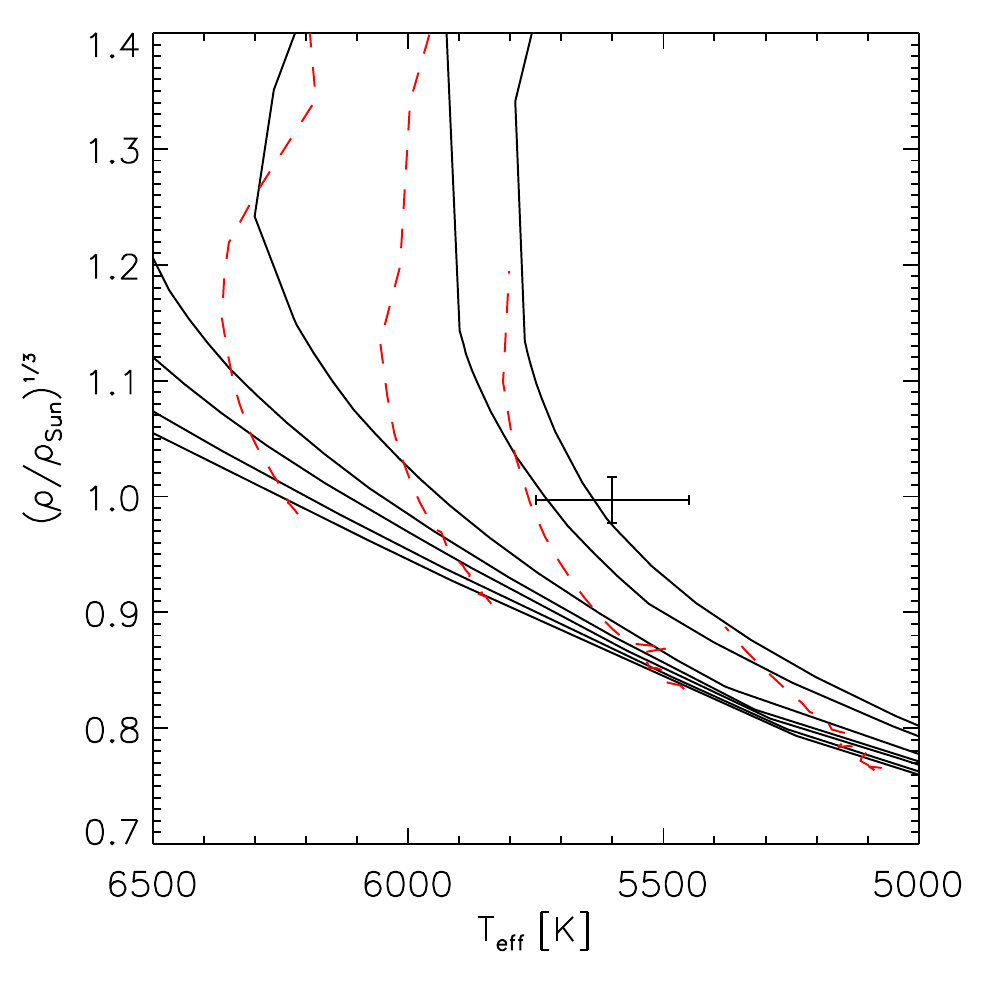}
}
\caption{The locations of WASP-42 and WASP-49 in the modified Hertzsprung--Russell diagram.}
\end{figure*}

\subsection{Combined Analysis}
\label{sec:comban}

The combined analysis of discovery and follow-up photometry and spectroscopic 
data was performed in several steps with the Markov Chain 
Monte Carlo (MCMC) code described in \citet{Gillon10a,Gillon12a}. 
The code makes use of the transit light curve models by \citet{Mandel02} and 
uses a Keplerian model to fit the radial-velocity measurements. For radial-velocity 
measurements obtained during transit, the prescription of the Rossiter--McLaughlin
effect provided by \citet{Gimenez06} is used. Various models for the photometric baseline
(e.g. 0th to 4th order polynomials with respect to time and external variables
such as pixel shifts) can be included in the fit of the transit light curves 
via minimization of the model coefficients at each MCMC step. 
The basic jump parameters are transit depth $dF$, impact parameter $b$,
transit duration $d$, epoch of mid-transit $T_{0}$, period $P$ and 
$K_{2} = K\sqrt{1-e^{2}}P^{1/3}$ (where $K$ and $e$ denote the radial velocity 
semi-amplitude and eccentricity, respectively). For these parameters we assume a uniform prior 
distribution. In order to take into account the stellar limb-darkening, we adopt a
quadratic law interpolating the coefficients tabulated by \citet{Claret11}. 
However, we chose their combination $c_{1} = 2u_{1} + u_{2}$ and $c_{2} = u_{1} - 2u_{2}$ 
as jump parameters as introduced by \citet{Holman06}. The physical basis of these parameters
leads us to assume a normal prior distribution with a width equal to the 
errors of these parameters. Where eccentricity is not set to $e = 0$,
it is included in the analysis via including the jump parameters 
$\sqrt{e}\cos\omega$ and $\sqrt{e}\sin\omega$ (where $\omega$ denotes the argument of
periastron).
We use the calibration technique devised by \citet{Enoch10} which uses a fit to a set of 
well studied main-sequence eclipsing binaries in order to infer the stellar mass and radius from the mean 
stellar density (measured directly from the transit light curves, \citet{Seager03}, 
temperature and metallicity. This technique is based on a similar relation 
by \citet{Torres10} where the stellar surface gravity is used in the place of the 
stellar density. A minimum of two chains is run in order to check their convergence
using the Gelman~\&~Rubin test \citep{Gelman92}.

As a first step, the photometric data obtained by the WASP-South cameras were
analyzed with the aim of finding a mean mid-transit time and period estimate from 
these long-term data. As the light curves from WASP contain many large outliers 
(i.e. points deviating by more than 10 times the transit depth), we discarded points deviating by 
more than $5 \sigma$ before running two MCMC chains of $10^{4}$ points each. Here the only 
parameters left to vary were the time of mid-transit and the period, while all other parameters 
were kept fixed at their approximate values. The period derived here was used as a starting point in the subsequent 
analysis of the high-precision follow-up data, while the mid-transit time was included as an 
extra constraint on the period.

Then, for each follow-up lightcurve we performed MCMC analyses including only the light curve
itself and the CORALIE radial velocity measurements. We tested various models for the 
photometric baseline, using chains of $10^{5}$ points. 
A second-order polynomial in time was assumed as a minimal baseline model to take care of stellar variability,
airmass and other time-dependent effects. More complicated models (i.e. polynomials up to the fourth 
degree with respect to time, position shifts, sky background, and FWHM) were tested with the 
Bayesian Information Criterion \citep{Schwarz78}. Only for the 
EulerCam light curve obtained on 19 January 2011 (WASP-49) was a more complicated model 
(a linear fit to position plus the quadratic fit to time) found to be justified. 

Finally, we performed a global analysis in order to find a single solution using all available 
follow-up data together with the mid-transit time obtained from the discovery photometry. The dependencies and orders
in the photometric baseline models were set to those found before. For each planet we ran three sets 
of two chains of $10^{5}$ points. From the first set, we determined correction factors
for our photometric errors based on the calculation of the red noise \citep{Pont06} 
as described in \citet{Winn08}. In the further analysis, the photometric errors were
multiplied by these factors. 
For the RV data, we determined ``jitter'' factors which serve to scale the measurement errors to the
standard deviation of the best-fit model. These factors are added quadratically to the RV errors and compensate
both instrumental and astrophysical effects (such as stellar activity) that are not included in the initial
error calculation. The values for the RV jitter are \mbox{0~m s$^{-1}$} 
(WASP-42, CORALIE), \mbox{2.4~m s$^{-1}$} (WASP-42, HARPS) and \mbox{10~m $s^{-1}$} (WASP-49, CORALIE). 

The second and third sets of chains were run with the adapted errors, first 
setting the eccentricity to $e=0$ and then leaving it free. For WASP-49, no significant
deviation from a circular orbit is detected \mbox{($e = 0.018_{-0.013}^{+0.023}$)}, and all parameters 
agree well within one sigma for circular and eccentric case. The eccentric solution for WASP-42 yields 
\mbox{$e=0.060_{-0.011}^{+0.013}$} and will be further discussed in Section \ref{sec:disc} while
the planetary and stellar parameters of both objects are presented in Table \ref{tab:par}. 

\begin{table*}
\centering                        
\begin{tabular}{p{3cm} p{3cm} p{3cm} p{3cm}}       
\hline\hline 
  & \multicolumn{2}{c}{WASP-42} \T & \multicolumn{1}{l}{WASP-49} \\    
\hline
 &  $e=0$ & $e \ne 0$  &  $e=0$ \T  \\   
\hline   
 \multicolumn{4}{l}{Jump parameters} \T  \\
\hline
 $ \Delta F = (R_{p}/R_{\ast})^{2}$ \T & $0.01650 \pm 0.00039$ &  ${0.01650 \pm 0.00037}$ & $0.01376\pm0.00038$ \\
 $ b' = a*\cos(i_{p})$ $[R_{\ast}]$ & $0.410^{+0.045}_{-0.055}$ &  ${0.418_{-0.056}^{+0.043}}$ & $0.745\pm0.014$ \\
 $ T_{14}$ [d] & $0.12042_{-0.00095}^{+0.00099}$ & ${0.12043 \pm 0.00093}$ & $0.08832\pm0.00080$ \\
 $ T_{[0]} - 2450000$ [HJD] & $5650.56723\pm0.00024$ & ${5650.56720\pm0.00023}$ & $5580.59436\pm0.00029$ \\
 $ P$ [d] & $4.9816877_{-0.0000073}^{+0.0000068}$ & ${4.9816872\pm0.0000073}$ & $2.7817387\pm0.0000056$ \\
 $ K_{2}$ [{\kmsa}] & $109.8 \pm 2.9 $ & ${110.4 \pm 2.9}$ & $79.9\pm3.3$ \\
 $ c_{1,\rm r'} $ & $1.271 \pm 0.056 $ & ${1.270 \pm 0.055}$ & $1.124_{-0.061}^{+0.058}$ \\
 $ c_{2,\rm r'} $ & $0.122\pm0.063 $ & ${0.120\pm0.065}$ &  $-0.130\pm0.048$ \\
 $ c_{1,\rm I+z'} $ &  $0.908\pm50.00 $ & ${0.907\pm-0.049}$ & $0.835\pm0.043$ \\
 $ c_{2,\rm I+z'} $ & $-0.234\pm0.033 $ & ${-0.234\pm0.032}$ & $-0.241\pm0.046$ \\
 $ c_{1,\rm z'} $ & - & - & $0.786\pm0.042$ \\
 $ c_{2,\rm z'} $ & - & - & $-0.257\pm0.027$ \\
\hline 
 \multicolumn{4}{l}{Deduced parameters} \T \\
\hline
 $ K $ [{\kms}] \T & $64.3 \pm 1.7$& ${64.8 \pm 1.7}$ & $56.8\pm2.4$ \\
 $ R_{p} $ [{\Rjup}] & $1.063\pm0.051 $ & ${1.080\pm0.057}$ & $1.115\pm0.047$ \\
 $ M_{p} $ [{\Mjup}] & $0.497\pm0.035 $ & ${0.500\pm0.035}$ & $0.378\pm0.027$ \\
 $ e $   & $0$ & ${0.060\pm0.013}$ & $0$ \\
 $ \omega $ [deg] & - & ${167\pm26}$ & -  \\
 $ a $ [AU] & $0.0547\pm0.0017$ & ${0.0548\pm0.0017}$ & $0.0379\pm0.0011$ \\
 $ a/R_{\ast} $ & $13.84\pm0.34 $ & ${13.65\pm0.46}$ & $8.35\pm0.16$  \\
 $ i_{p} $ [deg] & $88.30_{-0.23}^{+0.26} $ & ${88.25_{-0.23}^{+0.27}}$ & $84.89\pm0.19$  \\
 $ b_{tr} $  & $0.410_{-0.055}^{+0.045} $ &  ${0.411_{-0.054}^{+0.041}}$ & $0.745\pm0.014$  \\
 $ T_{occ} - 2450000$ [HJD]\tablefootmark{a} & $5658.03977\pm0.00024 $ & ${5652.889\pm0.035}$ & $5584.76697\pm0.00028$ \\
 $ \rho_{p} $ [{\rhojup}]  & $0.412_{-0.042}^{+0.049} $& ${0.397_{-0.047}^{+0.054}}$ & $0.273_{-0.026}^{+0.030}$ \\
 $ T_{eq} $ [K]\tablefootmark{b} & $988 \pm 31 $ & ${995 \pm 34}$ & $1369\pm39$  \\
 $ M_{\ast} $ [{\Msolar}]  & $0.881_{-0.081}^{+0.086} $ & ${0.884_{-0.080}^{+0.086}}$ & $0.938_{-0.076}^{+0.080}$  \\
 $ R_{\ast} $ [{\Rsolar}]  & $0.850\pm0.035 $ & ${0.863_{-0.034}^{+0.041}}$ & $0.976\pm0.034$  \\
 $ \rho_{\ast} $ [{\rhosun}] & $1.43\pm0.11 $ & ${1.37\pm0.14}$ & $1.0098\pm0.06$  \\
 $ u_{1,\rm r'} $ & $0.533 \pm 0.029 $ & ${0.532 \pm 0.029}$ & $0.424\pm0.029$  \\
 $ u_{2,\rm r'} $ & $0.205 \pm 0.025 $ & ${0.206 \pm 0.025}$ & $0.277\pm0.017$ \\
 $ u_{1,\rm I+z'} $ & $0.317 \pm 0.024 $ &  ${0.316 \pm 0.024}$ & $0.286\pm0.020$ \\
 $ u_{2,\rm I+z'} $ & $0.275 \pm 0.010 $ &  ${0.275 \pm 0.010}$ & $0.263\pm0.020$  \\
 $ u_{1,\rm z'} $ & - & - & $0.263 \pm 0.020$ \\
 $ u_{2,\rm z'} $ & - & - & $0.260 \pm 0.010$ \\ 
\hline
\multicolumn{4}{l}{Radial Velocity RMS} \T \\
\hline     
 CORALIE & 12.12 & 10.7 & 13.53  \T\\
 HARPS & 5.76 & 4.62 & - \\
 RMS/error HARPS & 1.00 &  0.80 & - \\
 RMS/error CORALIE & 0.96 & 0.85 & 1.56 \\
\hline
\end{tabular}
\caption{\label{tab:par}Planetary and stellar parameters for WASP-42 and WASP-49, as well as the radial-velocity RMS. For WASP-42 we show both the circular and the (preferred) eccentric model. 
\newline \tablefoottext{a}{Predicted}
\newline \tablefoottext{b}{Assuming A=0 and F=1.}}
                            
\end{table*}

\section{Discussion}
\label{sec:disc}

We announce the discovery of two new transiting extrasolar planets in the southern hemisphere from the WASP-South survey.

\subsection{WASP-42~b}

WASP-42~b is a 0.5~{\Mjup} planet orbiting the K1 star \mbox{2MASS 12515557--4204249} every 5 days. The host-star metallicity
is near the solar value. 

Based on the {\vsini} and stellar radius of WASP-42, we deduce a maximal rotation period of $P = 16.1 \pm 3.2$~d.
Using the \citet{Barnes07} relation, this gives an upper age limit of $\sim 0.84^{+0.51}_{-0.35}$~Gy from gyrochronology.
We derived a second estimate based on the $R'_{HK}$ activity indices \citep{Noyes84} from the HARPS spectra following 
the procedures of \citet{Lovis11}. We find $\log(R'_{HK}) = -4.9 \pm{0.7}$, 
indicative of a rather quiet star. Using the \citet{Mamajek08} relation we derive a rotation period of
$P = 40.1 \pm 4.7$~d and an age of $6.1\pm1.2$~Gy. We also interpolated the isochrones of \citet{Marigo08} 
using $z=0.021$ and compared them to the location of WASP-42 in the modified 
($(\rho_{\ast} / {\rhosun}) ^ {1/3}$ vs. {\teff}) Hertzsprung--Russell diagram (Figure \ref{fig:W42iso}). 
Here we used the temperature determined from the spectroscopic analysis together with the stellar density 
determined from the global analysis described in Section \ref{sec:comban}.
It should be noted that in the analysis of the spectra a degeneracy exists between macroturbulence {\mactrb} 
and stellar rotation {\vsini}. As the value for {\mactrb} is not directly measured but taken from tables, it might not be  
accurate for WASP-42. For example, a higher {\mactrb} would produce a lower {\vsini}, and thus a slower rotation and older 
gyrochronological age. Together with the absence of periodic brightness modulations (although we can not exclude
non-periodic photometric variations of a few milli-magnitudes), the evidence points towards an older age of several Gyrs
for WASP-42.

\begin{figure}
 \centering
 \includegraphics[width=\linewidth]{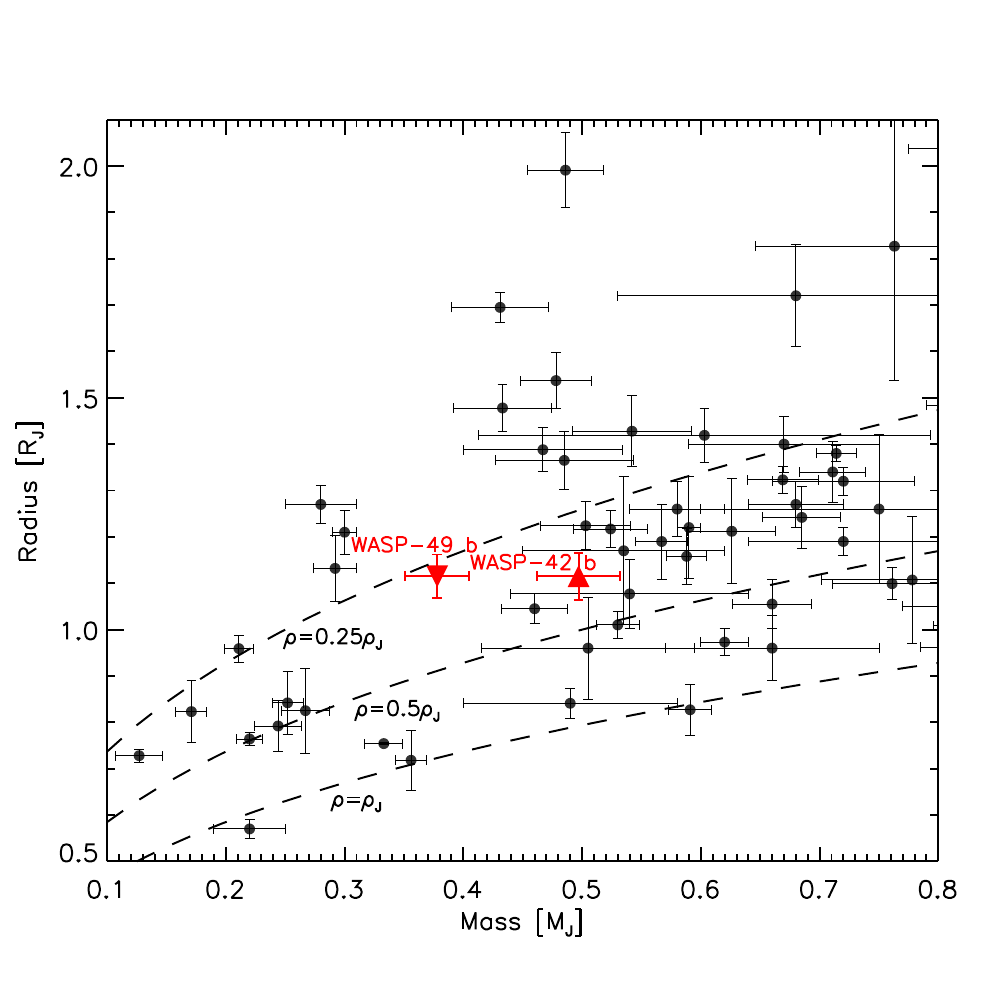}
    \caption{Mass--Radius diagram of transiting exoplanets. WASP-42~b and WASP-49~b are depicted using upward and downward
facing triangles, respectively. The dotted lines represent curves of equal density at (from top to bottom) $\rho=0.25$~{\rhojup}, $\rho=0.5$~{\rhojup}
and $\rho=$~{\rhojup}. The planet parameters were taken from \textit{www.exoplanet.eu}.}
\label{fig:MR}
\end{figure}

The planet's position in the mass--radius diagram (see Figure \ref{fig:MR}) is at the low-mass end 
of the Hot Jupiters. 
We used the tabulated radii of \citet{Fortney07} in order to check whether the planetary radius is 
within predictions using the values for \mbox{0.1 AU = 875 K}, and interpolating for a planetary mass of 0.5~{\Mjup}. 
Assuming the gyrochronologically determined age of 1 Gyr, we obtain a core mass of slightly above 10~{\Mearth}, while 
using values for 4.5 Gyrs would suggest a core mass close to 0~{\Mearth}. 
Compared to other planets of similar mass, WASP-42~b is no outlier in terms of density, but with $\rho_{p} = 0.40\pm0.05~{\rhojup}$ falls
into the well-populated region between 0.25~{\rhojup} and 0.5~{\rhojup}.
We also compared the planet's radius to the 
radius predicted from the fits describing the Saturn-mass population presented in \citet{Enoch12}.
Using Equation 8 of \citet{Enoch12} together with our uncertainties on stellar and planet parameters, 
we obtain a predicted radius of $1.31\pm0.03$~{\Rjup}, which is within the scatter around the model 
(Figure 14 in \citet{Enoch12}).

\begin{figure}
 \centering
 \includegraphics[width=\linewidth]{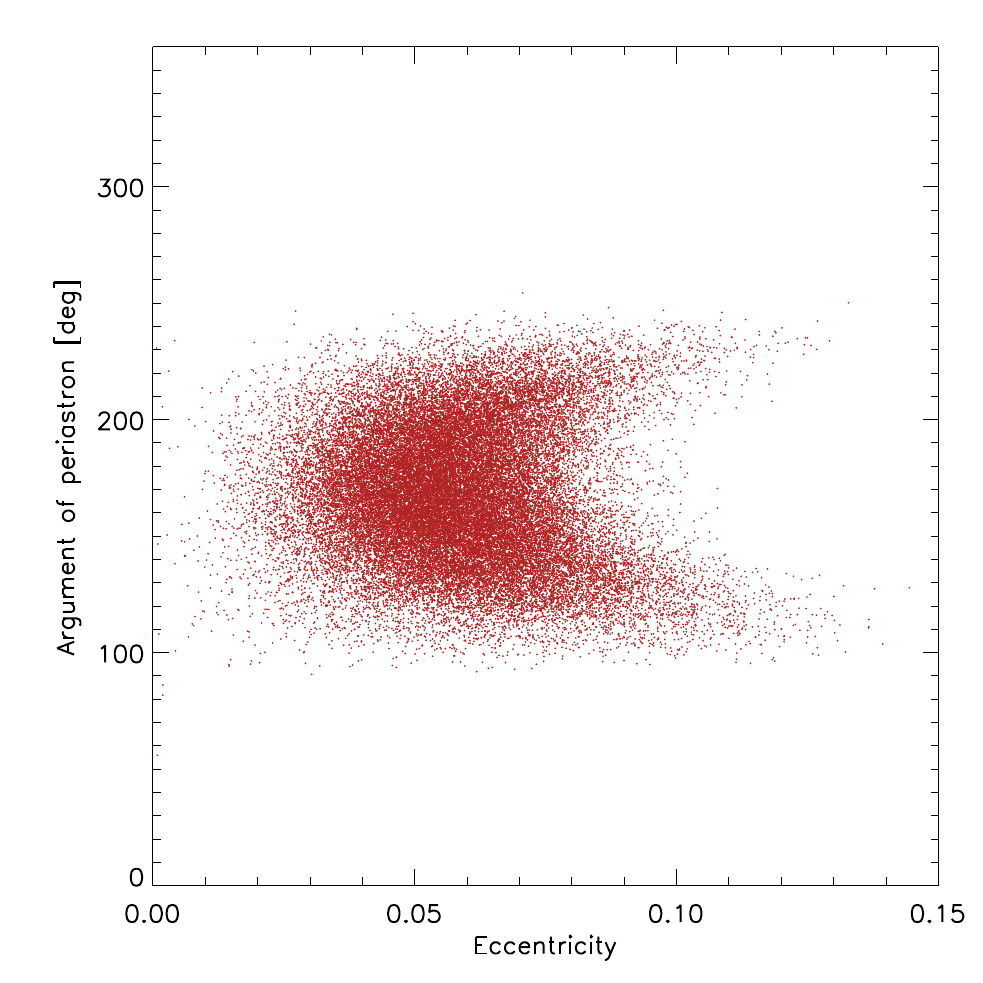}
    \caption{The distribution of eccentricity and argument of periastron for WASP-42~b obtained from the global analysis while
adding a jitter of 5~m~s$^{-1}$ to the HARPS and 15~m~s$^{-1}$ to the CORALIE data.}
\label{fig:W42oe}
\end{figure}

Adopting an eccentric orbit for WASP-42~b allows for a small ($0.060\pm0.013$) but non-zero eccentricity, while the 
argument of periastron is found to be $\omega = 167 \pm26$ deg. We checked any dependence of the derived eccentricity
on the jitter values assumed by re-running the analysis adding jitters of
3~m~s$^{-1}$ and 15~m~s$^{-1}$ quadratically to the CORALIE errors as well as a jitters of 5~m~s$^{-1}$ to the HARPS data. 
In all cases, the significance of the derived eccentricity remained above 3~$\sigma$. To illustrate, the distribution 
of the argument of periastron and eccentricity found from the
analysis including a jitter of 5~m~s$^{-1}$ for HARPS and 15~m~s$^{-1}$ for CORALIE is depicted in Figure \ref{fig:W42oe}.
In order to evaluate the robustness of this detection  
we computed the Bayes Factor \citep[e.g.][]{Carlin09} of the two models, resulting in a value of $B_{ec} = 2440$. This is 
a very strong indication that the eccentric model is a better representation of the data. To provide a second evaluation 
we compared the reduced chi-squared values $\chi_{red}^{2}$ of the radial-velocity models, for datasets including all 
data (CORALIE and HARPS), as well as all data but leaving out the radial-velocity sequence obtained on April 4 2011, for 
and the HARPS or CORALIE data separately. For the latter cases we re-ran 
the MCMC analyses and found eccentricities compatible with the one derived from the complete dataset. 
We also used the prescription of \citet{Lucy71} (LS test) in order to test each of the above for the significance of the 
circular solution. The results are summarized in Table \ref{tab:CHI}. In short, in all cases, the non-circular model gives a better fit 
to the data, and the values found for the eccentricity agree from all subsets. The LS test indicates a probability of
\mbox{99.5 \%} for the eccentric model to be required if the HARPS data are considered alone, and a \mbox{99 \%} probability if the 
entire dataset is considered, both above the \mbox{5 \%} limit suggested by \citet{Lucy71}. 
The CORALIE data alone do not justify the eccentric model. Considering that there are only 10 HARPS points 
outside the consecutive time series, more data would be highly desirable in order to secure the detection.
Based on these arguments, we conclude that there is considerable evidence for an eccentric orbit of WASP-42~b and we
present the eccentric as well as the circular solution in Table \ref{tab:par}.
 
The planet is located on the outer edge of the known Hot-Jupiter pileup, at a separation of 4.05 times the Roche limit, 
well above the 2 $R_{RL}$ cutoff identified by \citet{Ford06}, below which planets are thought to be unable to attain that location by the 
circularization of previously eccentric orbits. This means WASP-42~b is compatible with the evolutionary scenario 
described by, for example, \citet{Matsumura10}: formation at farther orbital separations, scattering via planet-planet or Kozai interactions
to an eccentric orbit, and subsequent tidal circularization. 

\begin{table*}[ht]
\centering                        
\begin{tabular}{ccccc} \hline \hline
Dataset & $e$ & $\chi_{red,~e\ne0}^{2}$ & $\chi_{red,~e=0}^{2}$ & LS test \tablefootmark{a} \T \\ \hline
HARPS + CORALIE & $0.060_{-0.011}^{+0.013}$ & $0.82\pm0.18$ & $1.33\pm0.22$  & 0.010 \T \\
HARPS + CORALIE (no RM sequence) &  $0.061_{-0.013}^{+0.014}$ & $0.87\pm 0.23$ & $1.53\pm0.30$ & 0.024 \\
CORALIE & $0.068_{-0.027}^{+0.029}$ & $1.13\pm0.34$ & $1.33\pm0.34$ & 0.24  \\
HARPS & $0.059_{-0.013}^{+0.016}$ & $0.90\pm0.26$ & $1.58\pm0.33$ & 0.005 \\
\\
\end{tabular}
\caption{Eccentricities, $\chi^{2}$ values and LS-test results of the eccentric and circular models for WASP-42~b found from all RV data and the datasets from the two instruments separately.
\newline \tablefoottext{a}{Probability of the circular model to be accurate, as defined by \citet{Lucy71}}.}
\label{tab:CHI}
\end{table*}

\subsection{WASP-49~b}

WASP-49~b is a near Saturn-mass \mbox{($ M = 0.38\pm0.03 {\Mjup}$)} planet in a 2.8-day orbit around the G6 star \mbox{2MASS 06042146--1657550}. 
In WASP-49, lack of lithium would suggest an age of several Gy
\citep{Sestito05} and the low {\vsini} implies a long rotation period of
around 50~d. The gyrochronological relation of \citet{Barnes07}
gives an age of $\sim 13^{+15}_{-8}$~Gy. Hence, we conclude that WASP-49 is a
relatively old main-sequence star. We performed an isochrone analysis the same way as for
WASP-42 with $z=0.011$ (Figure \ref{fig:W49iso}), finding an age above 10~Gyr, in good agreement with the age
found from gyrochronology.

In the period--mass plane WASP-49~b lies clearly below the bulk of Hot Jupiters and 
is nearly a twin of the well known HD149026 b \citep{Sato05}, a planet known for its high density. 
While the two planets have almost identical mass and period, they show very different 
radii and orbit two very different stars --
while WASP-49 is remarkable for having one of the lowest metallicities known for planet host stars 
($-$0.23 $\pm$ 0.07), HD149026 is one of the most metal-rich planet hosts. 
In contrast to HD149026 b, WASP-49~b has a low density ($0.27 \pm0.03$ {\rhojup}). 
WASP-49~b fills a gap between ~0.32 and ~0.42 {\Mjup} in the mass--radius plane (see Figure \ref{fig:MR}),
being the first low-density planet in this mass range.
Using the tabulated values from \citet{Fortney07} for system ages of $4.5$~Gyrs, an equilibrium
temperature of 1300 K and interpolating for a planet mass of 0.38 {\Mjup}, we find a predicted maximal planet radius of $1.08$ {\Rjup} 
(the case of a zero core mass). 
As with WASP-42~b, we employed Equation 8 of \citet{Enoch12} to obtain a predicted radius of $1.26\pm0.02$~{\Rjup}.
Again, this value is within the scatter around the fit (Figure 14 in \citet{Enoch12}).
We conclude that WASP-49~b, while larger than predicted from models, is not an outlier in the set of known planets.

There is no evidence for a non-zero eccentricity of WASP-49~b, and we determine a $3$-$\sigma$ upper limit of $e=0.09$.
WASP-49~b is separated from its host star by 2.4 times the Roche Limit, a rather low value which, together with the planet's old age, 
might indicate that orbital decay has occurred since the arrival of the planet at short orbital distances.

To summarize, WASP-42~b and WASP-49~b are two newly discovered close-in transiting planets with masses in the range 1--2 times
that of Saturn. While WASP-49~b is old, inflated and orbiting close to a metal-poor G6 star, WASP-42~b is orbiting a cooler K1 star 
at the outer edge of the well-known planet pile-up, probably on a slightly eccentric orbit.

\begin{acknowledgements}

WASP-South is hosted by the South African Astronomical Observatory 
and we are grateful for their ongoing support and assistance. 
Funding for WASP comes from consortium universities
and from the UK's Science and Technology Facilities Council.
TRAPPIST is funded by the Belgian Fund for Scientific  
Research (Fond National de la Recherche Scientifique, FNRS) under the  
grant FRFC 2.5.594.09.F, with the participation of the Swiss National  
Science Fundation (SNF). M. Gillon and E. Jehin are FNRS Research  
Associates. We would also like to thank the Geneva Stellar Variability
group, particularly Richard I. Anderson, for continuing flexible and 
short-notice time exchanges on the Euler-Swiss telescope and carefully 
conducted observations.

\end{acknowledgements}

\bibliographystyle{aa}
\bibliography{lendl_bbl}

\end{document}